\documentclass{pasj00}

\usepackage{epsfig}
\usepackage{natbib}
\usepackage{graphicx}

\begin{document}
\SetRunningHead{Jur\v{c}\'ak, J.}{The Analysis of Penumbral Fine Structure
Using an Advanced Inversion Technique}
\Received{2000/12/31}
\Accepted{2001/01/01}

\title{The Analysis of Penumbral Fine Structure Using an Advanced Inversion Technique}



%
\author{ Jan \textsc{Jur\v{c}\'ak}\altaffilmark{1,7}
    Luis \textsc{Bellot Rubio}\altaffilmark{2}
    Kiyoshi \textsc{Ichimoto}\altaffilmark{1}
    Yukio \textsc{Katsukawa}\altaffilmark{1}
    Bruce \textsc{Lites}\altaffilmark{3}
    Shin'ichi \textsc{Nagata}\altaffilmark{4}
    Toshifumi \textsc{Shimizu}\altaffilmark{5}
    Yoshinori \textsc{Suematsu}\altaffilmark{1}
    Theodore \textsc{Tarbell}\altaffilmark{6}
    Alan \textsc{Title}\altaffilmark{6}
    and
    Saku \textsc{Tsuneta}\altaffilmark{1}}
\altaffiltext{1}{National Astronomical Observatory of Japan, 2-21-1 Osawa,
Mitaka, Tokyo 181-8588, Japan} \altaffiltext{2}{Instituto de Astrof\'{\i}sica
de Andaluc\'{\i}a, Apdo. de Correos 3004, 18080 Granada, Spain}
\altaffiltext{3}{High Altitude Observatory, National Center for Atmospheric
Research, P.O. Box 3000, Boulder, CO 80307-3000, USA} \altaffiltext{4}{Hida
Observatory, Kyoto University, Kamitakara, Gifu 506-1314, Japan}
\altaffiltext{5}{Institute of Space and Astronautical Science, Japan Aerospace
Exploration Agency, Tokyo, Japan} \altaffiltext{6}{Lockheed Martin Solar and
Astrophysics Laboratory, Bldg. 252, 3251 Hanover St., Palo Alto, CA 94304, USA}
\altaffiltext{7}{Astronomical Institute of the Academy of Sciences, Fricova
298, 25165 Ond\v{r}ejov, Czech Republic}
\KeyWords{Sun: sunspots; methods: data analysis; techniques: polarimetric} 

\maketitle

\begin{abstract}
We present a method to study the penumbral fine structure using data obtained
by the spectropolarimeter onboard HINODE. For the first time, the penumbral
filaments can be considered as resolved in spectropolarimetric measurements.
This enables us to use inversion codes with only one-component model
atmospheres, and thus assign the obtained stratifications of plasma parameters
directly to the penumbral fine structure. This approach is applied to the
limb-side part of the penumbra in active region NOAA 10923. The preliminary
results show a clear dependence of the plasma parameters on continuum intensity
in the inner penumbra, i.e. weaker and horizontal magnetic field along with
increased line-of-sight velocity are found in the low layers of the bright
filaments. The results in the mid penumbra are ambiguous and future analyses
are necessary to unveil the magnetic field structure and other plasma
parameters there.
\end{abstract}

\section{Introduction}

Although the fine structure of penumbra has been studied for a long time, there
are only a few confirmed facts regarding the plasma properties and origin of
penumbral filaments. However, the global properties of the magnetic and
velocity structure of the penumbra are well known, i.e. the magnetic field
becomes weaker and more horizontal with increasing distance from the sunspot
umbra and the velocity field is composed mostly of horizontally oriented
Evershed flow, which points outward at photospheric layers  \citep[see the
review by][and references therein]{Solanki:2003}.

Differences in the plasma parameters of bright and dark filaments were reported
for the first time by \citet{Beckers:1969} who found stronger and more vertical
fields in dark filaments. With increasing spatial resolution came other
observations that confirmed the rapid changes of inclination and magnetic filed
strength on arcsec and sub-arcsec scales \citep[see][and references
therein]{Solanki:2003}.

However, the properties of the filamentary structure of the penumbra have been
derived indirectly from spectropolarimetric measurements (by means of
two-component model atmospheres) due to the relatively poor spatial resolution
attained by ground-based instruments \citep[see
e.g.][]{Bellot:2004,Bellot:2006,Borrero:2004,Borrero:2006}. These analyses
found weaker and more horizontal magnetic fields associated with increased
velocities, but these properties cannot be ascribed to specific intensity
structures due to the lack of spatial resolution.

Spectroscopic observations at 0.2$''$ resolution have revealed some properties
of bright penumbral filaments \citep{Bellot:2005} which are consistent with the
indirect analyses described above. Similar results were obtained also by
\citet{Langhans:2005} who analysed one-wavelength magnetograms at the same
resolution. A direct determination of the vector magnetic field of penumbral
filaments can be done for the first time using the spectropolarimetric data
obtained by HINODE Solar Optical Telescope (SOT).

In this paper, we describe the observations, data reduction, and the inversion
method used for a detailed analysis of the penumbral structure. Hopefully, our
results will help to distinguish between competitive models of the penumbra
\citep[see][for a review]{Bellot:2007}.

\section{Observations}

\begin{figure}[!th]
\centerline{\includegraphics[width=0.8\linewidth]{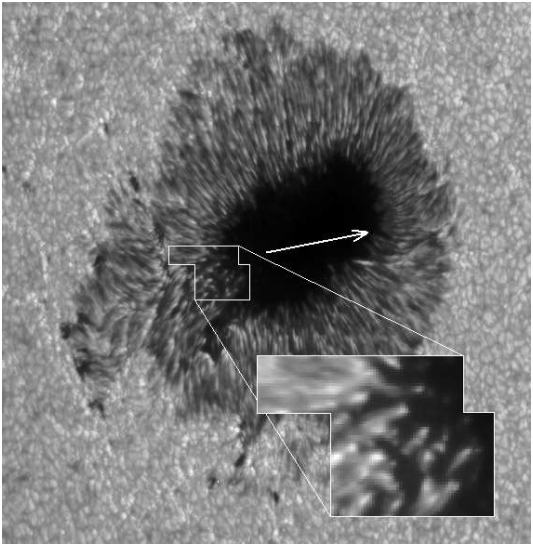}}
\caption{The map of continuum intensity reconstructed from the red wing of the
Fe~I~630.25~nm line. The area under analysis is highlighted and the arrow
points to disc centre.}\label{continuum}
\end{figure}

We analyse data obtained using the spectropolarimeter
\citep[SP;][]{Tarbell:2007} which is part of the Solar Optical Telescope
\citep[SOT;][]{Tsuneta:2007} onboard the HINODE satellite \citep{Kosugi:2007}.

The instrument observes the two iron lines Fe~I~630.15~nm (Land\'e factor
$g=1.67$) and Fe~I~630.25~nm ($g=2.5$). The diffraction limit of SOT is 0.3$''$
at 630~nm. The width of the spectrograph slit is equivalent to 0.16$''$
matching the pixel size of the CCD camera. The scanning steps of the
spectrograph slit is equivalent to 0.148$''$. The wavelength sampling of
2.15~pm is finer than the spectral resolution of 2.5~pm. The exposure time for
one slit position is 4.8~s, during which all Stokes profiles are acquired with
a noise level of $10^{-3} I_c$. The spatial resolution is slightly worse than
the diffraction limit of the telescope due to the aliasing induced by the CCD
pixel size, but it nevertheless reaches 0.32$''$.

The dark-field substraction, the flat-field division, the instrumental
polarisation correction, and other data reductions are made using the
calibration software developed by B. Lites. To prepare the data for the
inversion process, a calibration of wavelengths and a normalisation of the
Stokes profiles to the continuum intensity of the Harvard Smithsonian reference
atmosphere (HSRA) must be done. We perform this calibration for each slit
position separately using the pixels along the slit with weak polarisation
signal. The minimum of the Fe~I~630.15~nm Stokes $I$ profile averaged over
these pixels is used for the definition of zero velocity and the continuum
intensity of this profile is normalised to HSRA as unity (times the appropriate
limb-darkening factor). Following this approach, the photospheric 5~minute
oscillations are suppressed to the extent that we do not see any oscillations
in the resulting maps of line-of-sight (LOS) velocity.

\begin{figure}[!th]
\centerline{\includegraphics[width=\linewidth]{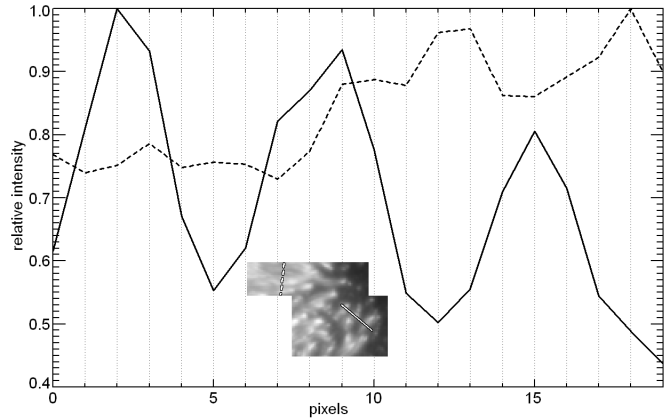}} \caption{Two
cuts through the filamentary structure of the penumbra are marked in the
intensity map in the lower part of the figure. The continuum intensity along
these cuts are shown in the plot where the solid and dashed lines correspond to
the cuts through the inner and mid penumbra respectively.}\label{diameter}
\end{figure}

On November 10, 2006, the spot in AR~10923 located at heliocentric position
49$^{\circ}$~E and 6$^{\circ}$~S (heliocentric angle $\mu=0.65$) was observed.
The observed region is shown in figure~\ref{continuum}, where the highlighted
area in size of $14\times10$~arcsec is analysed. The arrow points to the disc
centre.

In figure~\ref{diameter} we demonstrate that the penumbral filaments are
resolved by the SP observations for the first time. There are two cuts through
the filamentary structure shown in the lower part of the figure, one in the
inner penumbra (solid line) and one in the mid penumbra (dashed line). We plot
the continuum intensity along these cuts. Usually, the brightenings and
darkenings occupy at least two pixels (0.32$''$, i.e. comparable to the spatial
resolution of the instrument). Even if the intensity structures are smaller,
they still occupy the majority of the resolution element and thus dominate the
line forming process.

\section{Inversion Method}

We use a modification of the inversion code SIR (Stokes Inversion based on
Response functions) developed at the Instituto de Astrof\'{\i}sica de Canarias
\citep{Cobo:1992}. This code works under the assumption of local
thermodynamical equilibrium and hydrostatic equilibrium. See the survey studies
by \citet{Cobo:1998} or \citet{Jurcak:disertace}\footnote{this can be
downloaded at\\ http://www.asu.cas.cz/$\sim$sdsa/jurcak\_en.html}. The
inversion code synthesises the Stokes profiles coming from an initial model
atmosphere and compares them to the observed ones. Using a least-square
Marquardt's algorithm, the atmospheric model is modified until the difference
between the observed and synthetic Stokes profiles (merit function, $\chi^2$)
is minimised.

The modification of the SIR code was made by L.\ Bellot Rubio to allow for
gaussian perturbation in the stratifications of plasma parameters along the
line of sight, somewhere in the line forming region.  This modified code
\citep[][hereafter called SIR/GAUSS]{Bellot:2003} uses by definition a
two-component model atmosphere. The first component represents a background
atmosphere with simple stratifications of plasma parameters. The second
component is equal to the background atmosphere, but with gaussian perturbation
superposed to it, i.e. the second component can be exactly the same as the
first one at some heights depending on the width and position of the gaussian.

\begin{figure}[!t]
\centerline{\includegraphics[width=\linewidth]{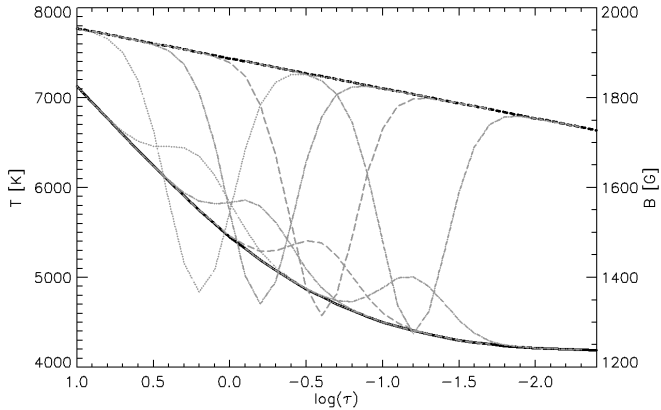}} \caption{The
background stratifications of temperature (solid black line) and magnetic field
strength (dashed black line). The various styles of gray lines represents the
gaussian perturbations added to the stratifications of the background model of
atmosphere at four different heights.}\label{invcode}
\end{figure}

Figure~\ref{invcode} shows examples of the background temperature
stratification (solid black line) and the magnetic field strength
stratification (dashed black line) where the gaussian perturbations are
represented by various styles of gray lines. Since we argue that the fine
structure of the penumbra is resolved in these SP observations, we set the
filling factor of the background component to $0.1~\%$ and thus the inversion
code works in practise with a simple one-component model atmosphere.

The original SIR code converges to the same resulting model of atmosphere
almost independently of the initial model. However, the modified code cannot
change the stratifications of plasma parameters independently as the original
one since there are ties between the parameters induced by the gaussian
perturbation which must have the same width and position for all of them.
Therefore, the resulting model of atmosphere depends on the starting one,
especially on the initial position of the gaussian perturbation. The Stokes
profiles observed in the analysed region are fitted using four different
starting positions for the gaussian perturbation (see figure~\ref{invcode}).
The merit functions of the resulting models of atmosphere are compared at each
pixel and the best solution is used to create maps of plasma parameters.

The starting positions of the gaussian perturbations are $\log (\tau) =
0.2\textrm{,} -0.2\textrm{,} -0.6\textrm{, and} -1.2$. Except for the height of
the gaussians, the other parameters of the initial guess models are the same:
HWHM, $\sigma$, of the gaussian function is 0.5 in units of the logarithm of
optical depth, the amplitude of the perturbation is $+$800~K for the
temperature, $-500$~G for the magnetic field strength, $+$3~km~s$^{-1}$ for the
line-of-sight velocity, $-30^\circ$ for the field inclination, and $-5^\circ$
for the field azimuth. The background stratifications are the same in all four
cases. In total, the inversion retrieves 13 free parameters.

The initial amplitudes of the gaussian perturbation have been chosen on the
basis of two-component inversions of the penumbral fine structure
\citep[e.g.][]{Bellot:2004, Borrero:2006} and because such perturbations can
represent currently available theoretical models of the penumbra: the moving
flux tube model \citep{Schlichenmaier:1998} and the field-free gap model
\citep{Scharmer:2006}. These two models have similar characteristics from an
observational point of view, that is, weaker and rather horizontal fields
embedded in a strong and less inclined background field. The moving flux tube
model predicts that the weak fields are associated with increased velocities.
Gaussian perturbations with initial amplitudes such as the ones used here can
simulate the top of field-free gaps or low-lying penumbral flux tubes if they
are positioned around the continuum forming layer ($\log (\tau)=0.2$ or
$-0.2$), or penumbral flux tubes if they are located higher in the line forming
region.

\begin{figure*}[!th]
\centerline{\includegraphics[width=0.9\linewidth]{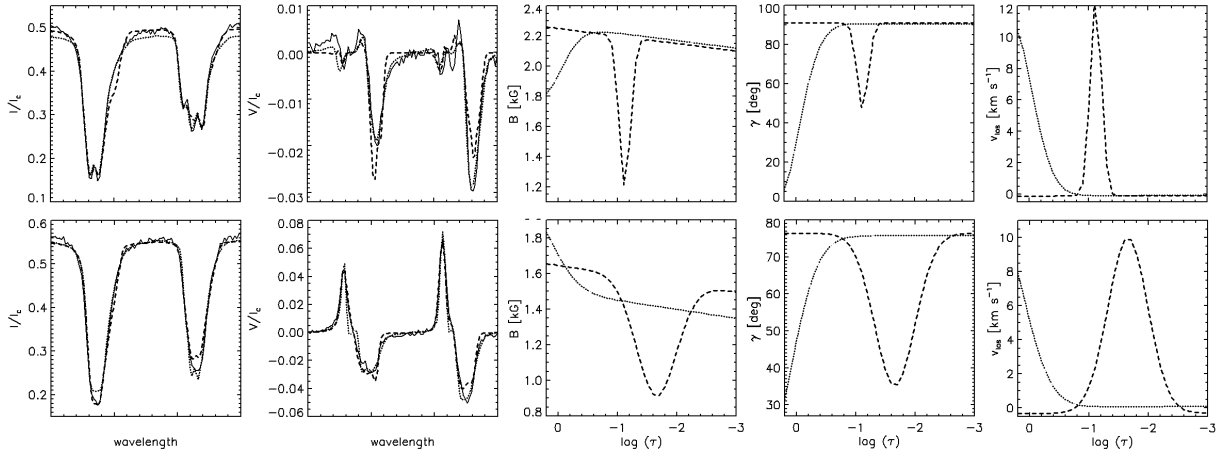}} \caption{
Observed Stokes $I$ and $V$ profiles (solid lines) plotted together with the
synthetic profiles and stratifications of magnetic field strength, inclination,
and LOS velocity obtained using the SIR/GAUSS code with starting position of
the gaussian perturbation at $\log (\tau) = 0.2$ (dotted lines) and $\log
(\tau) = -1.2$ (dashed lines). The upper row shows an example of the profiles
observed in the inner penumbra, where the mid penumbra is represented by the
profiles displayed in the lower row.}\label{pixels}
\end{figure*}
\begin{figure*}[!t]
\centerline{\includegraphics[width=0.9\linewidth]{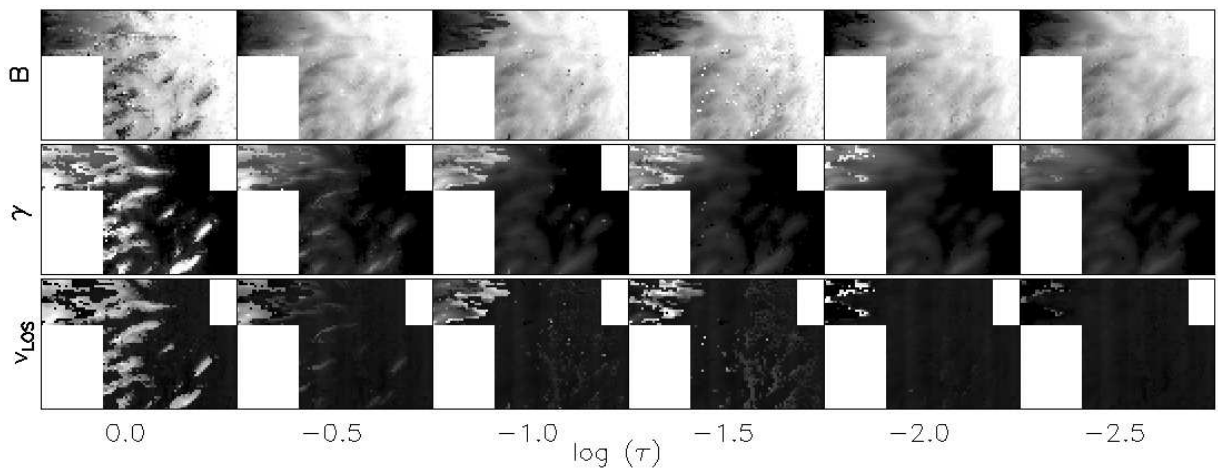}}
\caption{Resulting maps of magnetic field strength, inclination, and
LOS~velocity (from top to bottom) at different heights in the atmosphere. The
colour codes for displayed plasma parameters are the same as in
figure~\ref{cut}, see the colour bars there.}\label{mapy}
\end{figure*}

\section{Penumbral Fine Structure}

As only a small part of the penumbra is analysed and the studied area does not
cover the whole width of the penumbra, no general conclusions about the
penumbral structure are made here. We selected this region because lots of
clearly defined bright filaments enter the umbra, enabling us to study their
plasma properties with high precision using the method described above.
Moreover, the Stokes $V$ profiles observed in the limb-side penumbra at a given
position on the solar disc are highly asymmetric. The asymmetries can be
explained only with gradients of velocity and magnetic field in the line
forming region and thus the presence of gaussian perturbations in the
stratifications of these plasma parameters is necessary because the gradients
in the background component are too small to produce such asymmetric profiles.

Figure~\ref{pixels} shows two examples of observed profiles. The upper row
displays Stokes $I$ and $V$ profiles observed in the inner penumbra (solid
lines) together with the synthetic profiles and stratifications of magnetic
field strength, inclination, and LOS velocity obtained using the SIR/GAUSS code
with starting position of the gaussian perturbation at $\log (\tau) = 0.2$
(dotted lines) and $\log (\tau) = -1.2$ (dashed lines). The same applies for
the lower row, where a pixel from the mid penumbra is studied.

The inclination is evaluated with respect to the line of sight in these plots.
We can see that the inclination of the background field is in both cases close
to $90^\circ$ and thus the $V$ profiles generated by this component are weak.
In the inner penumbra the background component has opposite polarity with
respect to the gaussian component (it is slightly larger than $90^\circ$) and
that together with the velocity difference are the main reasons for the
asymmetric $V$ profile. The polarity of the background and gaussian components
are the same in the mid penumbra, therefore we do not see any $V$ lobe
cancellation and the asymmetry is caused only by the difference in LOS
velocities.

The solution in the inner penumbra is easy to interpret as the best fits are
always obtained using gaussian perturbation deep in the atmosphere. In the
example shown in the upper row of figure~\ref{pixels}, the merit function is
$12.3$ for the solution obtained with gaussian perturbation low in the
atmosphere compared to $\chi^2=22.6$ for that obtained with the gaussian
located higher in the line forming region.

On the other hand, the correct solution in the mid penumbra is much harder to
find. In the example of figure~\ref{pixels}, the fit obtained using the initial
perturbation high in the atmosphere is better ($\chi^2$=18.2 for
$\log(\tau)=-1.2$ vs. $\chi^2$=25.9 for $\log(\tau)=0.2$), although the shown
fits to the Stokes $I$ and $V$ profiles look like of a similar quality.
Generally, the variations in merit functions obtained with different starting
heights for the gaussian perturbations are not significantly different in the
mid penumbra and thus it is difficult to decide which solution is closer to
reality. Moreover, the fits are generally worse in the mid penumbra, where
$\langle\chi^2\rangle=33$, than in the inner penumbra, where
$\langle\chi^2\rangle=19$.

\begin{figure*}[!th]
\centerline{\includegraphics[width=0.9\linewidth]{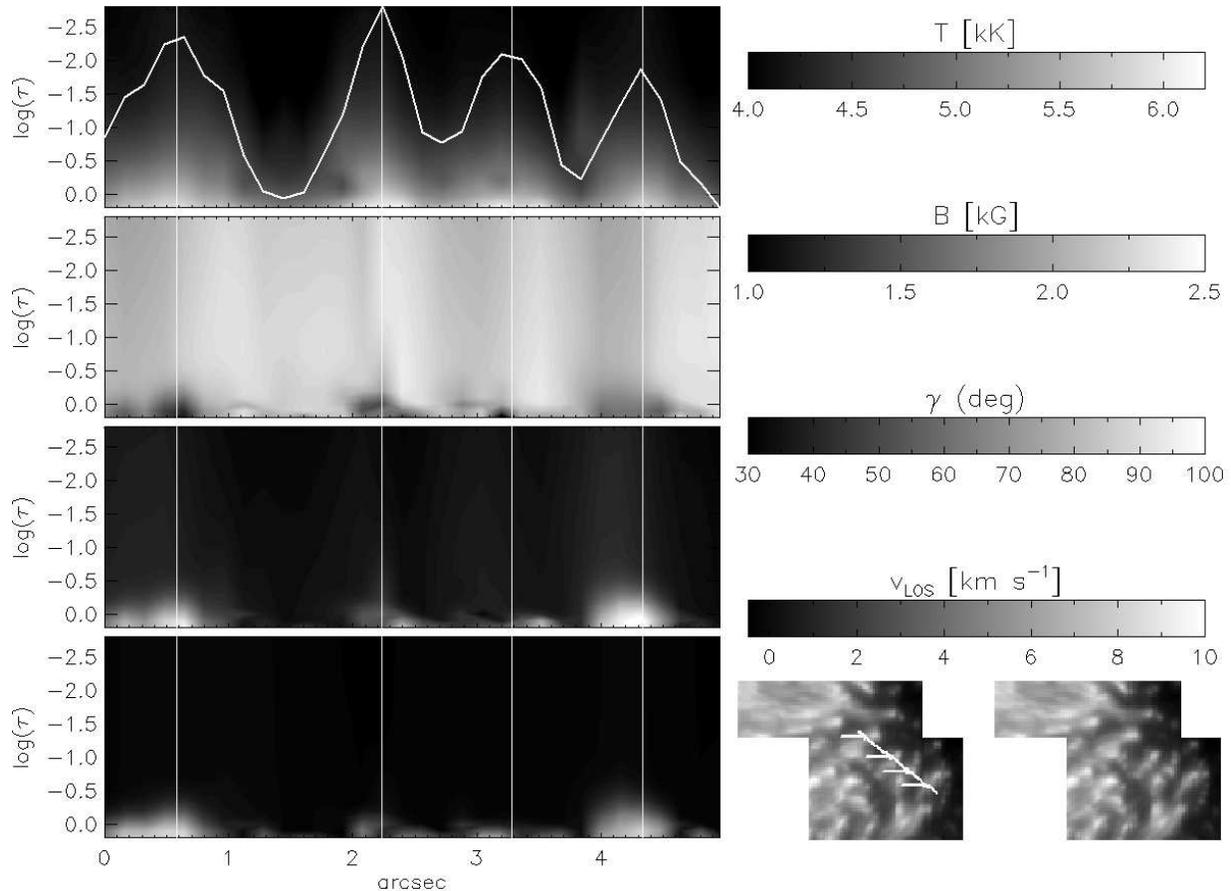}}
\caption{Vertical cut through four bright structures in the inner penumbra is
presented. The position of the cut is indicated in the continuum intensity map
in the lower left part of the figure. The left part of the figure shows the
stratifications of temperature, magnetic field strength, inclination, and
LOS~velocity (from top to bottom) along this cut, where the right part of the
figure displays the colour bars of these plasma parameters. The white line in
the temperature stratification map corresponds to the continuum intensity along
the cut. The vertical lines in the maps of stratifications correspond to the
arrows in the continuum intensity map.}\label{cut}
\end{figure*}

In figure~\ref{mapy} we show the resulting maps of magnetic field strength
(upper row), inclination (middle row), and LOS~velocity (lower row) at six
optical depths. These maps are created from the best solutions which were
obtained at each point (selected on the basis of merit function).

This figure clearly shows the discontinuities in the values of plasma
parameters in the mid penumbra (upper left part of the analysed region). The
problem is that solutions with gaussian perturbation located low and high in
the line forming region reproduce the observed Stokes profiles with similar
quality and do have different stratifications of plasma parameters (although
the resulting stratifications of the background component are similar). At some
areas in the mid penumbra we found weaker and more horizontal magnetic field
together with increased LOS~velocities high in the atmosphere (around
$\log(\tau)=-1.5$). A typical example of such a solution is the pixel presented
in the lower part of figure~\ref{pixels}, dashed line. At other areas the best
fit corresponds to the location of the gaussian perturbation low in the
atmosphere, i.e. the field is more horizontal and the LOS~velocities are
increased there. However, for this type of solution the magnetic field strength
is either slightly larger or remains basically the same as in the background
component (i.e. the amplitude of the gaussian perturbation in field strength is
small, see the dotted line in the lower part of figure~\ref{pixels} as an
example).

As can be seen, the solution is clear in the inner penumbra, where the best
fits correspond to the gaussian perturbation low in the atmosphere. The bright
filaments protruding into the umbra have lower magnetic field strengths, more
horizontal fields, and significantly larger LOS~velocities. Similar conclusions
can be made on the basis of figure~\ref{cut}, where the stratifications of
various plasma parameters are shown along a cut that cross four bright
structures in the inner penumbra (see the caption of this figure for details).
In figures~\ref{mapy} and~\ref{cut}, the inclination has been transformed to
the local reference frame, i.e. it is measured from the local normal line
pointing downwards.

In the case of the magnetic field inclination and LOS~velocities, the
background stratifications are not significantly influenced by the presence of
bright structures and reach similar values as in the surrounding umbra, i.e.
the filamentary structure almost disappears higher in the atmosphere. In the
case of the stratifications of the magnetic field strength, the filamentary
structure is sharpest low in the atmosphere but do not disappear with height.
This could be an artefact of the inversion technique, but it remains to be
tested.

In the upper map of figure~\ref{cut} we show the temperature along the cut and
the white line representing the continuum intensity along the cut. We can
clearly see that the bright structures have naturally higher temperatures low
in the atmosphere. The increase of the temperature in such structures is
imperative, but the absolute values of the temperature enhancement cannot be
always trusted as we occasionally do not fit the continuum of Stokes $I$
profile correctly (mostly because the macro-turbulence, the helping parameter
for the line broadening, is set to zero and thus the increase of temperature at
lower layers would result into better fit of the continuum intensity but worse
fit of the line wings).

Here we make the first rough estimates of the difference between the values of
plasma parameters in the bright filaments and the surrounding umbra. The
parameters given below correspond to optical depth $\log(\tau)=-0.2$, where the
results of the inversion are more reliable than at optical depth unity layer
itself. The uncertainties in the plasma parameters will not be discussed here,
but are much smaller than the differences we find.

The estimates are made from the bright filaments whose cut is analysed in
figure~\ref{cut}. In the case of the magnetic field strength we find a decrease
of approximately 600~G (1700~G in the bright filaments compared to 2300~G in
the surrounding umbra). This value is similar to those deduced from the
two-component inversions performed by \citet{Bellot:2004} and
\citet{Borrero:2006}, i.e. 500~G and 1500~G respectively. In the inner
penumbra, these authors found magnetic field inclination around 20$^\circ$ and
60$^\circ$ in the background and the flux-tube component, respectively. The
values in the surrounding umbra should be comparable to those of the background
atmosphere in two-component inversions, and indeed we find values around
30$^\circ$. However, the field inclinations in bright filaments are around
90$^\circ$ in our case, larger than those derived from two-component
inversions. As expected, the LOS~velocities are close to zero in the umbra
surrounding the bright filaments and reach values around 4~km~s$^{-1}$ in them.
These large velocities are again compatible with the results of
\citet{Bellot:2004} and \citet{Borrero:2006}, who found values of 5 and
3~km~s$^{-1}$ in the flux-tube component for spots at similar heliocentric
angles (40$^\circ$, resp 37$^\circ$). The most important point is that we can
directly ascribe these properties to the bright penumbral filaments. That was
not possible with the results of the two-component inversions mentioned above.

\section{Conclusions}

We have introduced a method to analyse the data obtained by SP onboard HINODE
using an advanced inversion technique. We have applied this method to the
limb-side penumbra of Active Region 10923, observed on November 10, 2006. This
allows us to directly obtain the structure of the magnetic field vector
together with the velocity configuration of the penumbral fine structure. This
is possible for the first time and our preliminary results in the inner
penumbra confirm previous results coming from two-component inversions of
visible and near-infrared lines. However, inversion techniques have so far been
unable to relate the magnetic structures represented by the two components with
intensity structures. This had to be done with the help of spectroscopic data
or magnetograms which have better spatial resolution.

Using SP data from HINODE SOT, we find that the magnetic field in the bright
filaments is weaker by some 600 G compared with the surrounding umbra at
optical depth $\log(\tau)=-0.2$. The field is horizontal in those layers and
the LOS velocity reaches values around 4~km~s$^{-1}$. The differences between
the plasma parameters in the bright filaments and the surrounding umbra quickly
decrease with height and disappear around $\log(\tau)=-0.5$. This means that
the bright filaments are structures located deep in the atmosphere in the inner
penumbra and that the background magnetic field closes above them. To some
extent, such a result is similar to that found in light bridges
\citep{Jurcak:2006}.

This could support the theory of a field-free gappy penumbra
\citep{Scharmer:2006} which hypothesises that bright penumbral filaments and
light bridges are the representations of the same phenomena (top of field-free
gaps) of different magnitude. However, our results could be also explained by
flux tubes located around optical depth unity in the inner penumbra, as
predicted by the moving tube model of \citet{Schlichenmaier:1998}.

The advantage of this model is that strong Evershed flows channeled by the
tubes arise naturally from the calculations, whereas the gappy model does not
offer any explanation for the Evershed flow. One of the argument against the
moving tube model is the large predicted outflow velocity around
13~km~s$^{-1}$; here we find similar magnitudes of velocities in the mid
penumbra. For example 11~km~s$^{-1}$, if we compute the outflow velocity as
$v_{\textrm{\small{LOS}}}/\cos(\gamma_{\textrm{\small{LOS}}})$ and take the
value of LOS~velocity of 7~km~s$^{-1}$ and LOS~inclination of 50~deg (selected
on the basis of the pixel from mid penumbra shown in figure~\ref{pixels},
dashed lines, where we do not use the peak values as they are probably
overestimated).

The results in the mid penumbra are more difficult to interpret. There are two
possible solutions which fit the observed Stokes profiles with similar quality.
According to the rising flux tube model, the tubes should be positioned higher
in the atmosphere in the outer penumbra and thus a more careful and detailed
analysis of this region is needed to test the realism of the two most discussed
models of the penumbral fine structure.
\\
\\
This work has been enabled thanks to the funding provided by the Japan Society
for the Promotion of Science. Hinode is a Japanese mission developed and
launched by ISAS/JAXA, with NAOJ as domestic partner and NASA and STFC (UK) as
international partners. It is operated by these agencies in co-operation with
ESA and NSC (Norway). The computations were partly carried out at the NAOJ
Hinode Science Center, which is supported by the Grant-in-Aid for Creative
Scientic Research The Basic Study of Space Weather Prediction from MEXT, Japan
(Head Investigator: K. Shibata), generous donations from Sun Microsystems, and
NAOJ internal funding.


\begin{thebibliography}{}

\bibitem[{{Beckers} \& {Schr{\" o}ter}(1969)}]{Beckers:1969}
{Beckers}, J.~M. \& {Schr{\" o}ter}, E.~H. 1969, Sol. Phys., 10, 384

\bibitem[{{Bellot Rubio}(2003)}]{Bellot:2003}
{Bellot Rubio}, L.~R. 2003, in Astronomical Society of the Pacific Conference
  Series, Vol. 307, Astronomical Society of the Pacific Conference Series, ed.
  J.~{Trujillo-Bueno} \& J.~{Sanchez Almeida}, 301

\bibitem[{{Bellot Rubio} {et~al.}(2004){Bellot Rubio}, {Balthasar}, \&
  {Collados}}]{Bellot:2004}
{Bellot Rubio}, L.~R., {Balthasar}, H., \& {Collados}, M. 2004, A\&A, 427, 319

\bibitem[{{Bellot Rubio} {et~al.}(2005){Bellot Rubio}, {Langhans}, \&
  {Schlichenmaier}}]{Bellot:2005}
{Bellot Rubio}, L.~R., {Langhans}, K., \& {Schlichenmaier}, R. 2005, \aap, 443,
  L7

\bibitem[{{Bellot Rubio} {et~al.}(2006){Bellot Rubio}, {Schlichenmaier}, \&
  {Tritschler}}]{Bellot:2006}
{Bellot Rubio}, L.~R., {Schlichenmaier}, R., \& {Tritschler}, A. 2006, A\&A,
  453, 1117

\bibitem[{{BellotRubio}(2007)}]{Bellot:2007}
{BellotRubio}, L.~R. 2007, in Highlights of Spanish Astrophysics IV, eds. F.
  Figueras, J.M. Girart, M. Hernanz, and C. Jordi (Springer), in press
  [astroph/0611471]

\bibitem[{{Borrero} {et~al.}(2004){Borrero}, {Solanki}, {Bellot Rubio}, {Lagg},
  \& {Mathew}}]{Borrero:2004}
{Borrero}, J.~M., {Solanki}, S.~K., {Bellot Rubio}, L.~R., {Lagg}, A., \&
  {Mathew}, S.~K. 2004, A\&A, 422, 1093

\bibitem[{{Borrero} {et~al.}(2006){Borrero}, {Solanki}, {Lagg},
  {Socas-Navarro}, \& {Lites}}]{Borrero:2006}
{Borrero}, J.~M., {Solanki}, S.~K., {Lagg}, A., {Socas-Navarro}, H., \&
  {Lites}, B. 2006, A\&A, 450, 383

\bibitem[{{Jur{\v c}{\'a}k} {et~al.}(2006){Jur{\v c}{\'a}k}, {Mart{\'{\i}}nez
  Pillet}, \& {Sobotka}}]{Jurcak:2006}
{Jur{\v c}{\'a}k}, J., {Mart{\'{\i}}nez Pillet}, V., \& {Sobotka}, M. 2006,
  A\&A, 453, 1079

\bibitem[{{Jur\v{c}\'ak}(2006)}]{Jurcak:disertace}
{Jur\v{c}\'ak}, J. 2006, {Two dimensional spectropolarimetry of a sunspot,
  doctoral thesis, Charles University, Prague}

\bibitem[{{Kosugi}(2007)}]{Kosugi:2007}
{Kosugi}, T., et al. 2007, Sol. Phys., to be submited

\bibitem[{{Langhans} {et~al.}(2005){Langhans}, {Scharmer}, {Kiselman},
  {L{\"o}fdahl}, \& {Berger}}]{Langhans:2005}
{Langhans}, K., {Scharmer}, G.~B., {Kiselman}, D., {L{\"o}fdahl}, M.~G., \&
  {Berger}, T.~E. 2005, A\&A, 436, 1087

\bibitem[{{Ruiz Cobo}(1998)}]{Cobo:1998}
{Ruiz Cobo}, B. 1998, Ap\&SS, 263, 331

\bibitem[{{Ruiz Cobo} \& {del Toro Iniesta}(1992)}]{Cobo:1992}
{Ruiz Cobo}, B. \& {del Toro Iniesta}, J.~C. 1992, ApJ, 398, 375

\bibitem[{{Scharmer} \& {Spruit}(2006)}]{Scharmer:2006}
{Scharmer}, G.~B. \& {Spruit}, H.~C. 2006, A\&A, 460, 605

\bibitem[{{Schlichenmaier} {et~al.}(1998){Schlichenmaier}, {Jahn}, \&
  {Schmidt}}]{Schlichenmaier:1998}
{Schlichenmaier}, R., {Jahn}, K., \& {Schmidt}, H.~U. 1998, A\&A, 337, 897

\bibitem[{{Solanki}(2003)}]{Solanki:2003}
{Solanki}, S.~K. 2003, A\&A Rev., 11, 153

\bibitem[{{Tarbell}(2007)}]{Tarbell:2007}
{Tarbell}, T., et al. 2007, Sol. Phys., to be submited

\bibitem[{{Tsuneta}(2007)}]{Tsuneta:2007}
{Tsuneta}, S., et al. 2007, Sol. Phys., to be submited

\end{thebibliography}
\end{document}